\documentclass[final, twocolumn]{elsarticle}

\usepackage{amsmath}
\usepackage{amssymb}
\usepackage{hyperref}
\usepackage{svg}
\graphicspath{Figures/}

\usepackage{tikz}

\usetikzlibrary{shapes,arrows, decorations.text}

\tikzstyle{decision} = [diamond, draw, fill=blue!20, 
    text width=4.5em, text badly centered, inner sep=0pt]
\tikzstyle{block} = [rectangle, draw, fill=blue!20, 
    text width=9em, text centered, rounded corners, minimum height=4em]
\tikzstyle{start} = [rectangle, draw, fill=green!40, 
    text width=9em, text centered, minimum height=4em]
\tikzstyle{line} = [draw, -latex']
\tikzstyle{cloud} = [draw, ellipse,fill=red!20, minimum height=2em, 
    text width=10.5em, text centered]

\usepackage{dcolumn}
\usepackage{bm}
\usepackage[caption=false]{subfig}

\newcommand{\eV}{\; \text{eV}}
\newcommand{\MeV}{\; \text{MeV}}
\newcommand{\GeV}{\; \text{GeV}}
\newcommand{\MHz}{\; \text{MHz}}
\newcommand{\GHz}{\; \text{GHz}}
\newcommand{\Xmax}{\; X_{\text{max}}}
\newcommand{\gcm}{\; \text{g}/\text{cm}^2}

\newcommand{\m}{\; \text{m}}

\newcommand{\textMHz}{$\text{MHz}$}

\newcommand{\textXmax}{$X_{\text{max}}$}
\newcommand{\textgcm}{$\text{g} / \text{cm}^2$}

\newcommand{\slice}{\text{slice}}
\newcommand{\ant}{\text{ant}}
\newcommand{\template}{\text{template}}
\newcommand{\target}{\text{target}}
\newcommand{\origin}{\text{origin}}

\journal{Astroparticle Physics}

\begin{document}

\begin{frontmatter}
    
\title{Proof of principle for template synthesis approach \\ for the radio emission from vertical extensive air showers}

\author[vub]{Mitja Desmet}
    \ead{mitja.desmet@vub.be}
\author[vub]{Stijn Buitink}
\author[vub,kit]{Tim Huege}
\author[kit]{David Butler}
\author[kit]{Ralph Engel}
\author[vub,rug]{Olaf Scholten}

\affiliation[vub]{
    organization={Astrophysical Institute, Vrije Universiteit Brussel (VUB)},
    addressline={Pleinlaan 2},
    city={Brussels},
    postcode={1050},
    country={Belgium}
}

\affiliation[kit]{
    organization={Institute for Astroparticle Physics, Karlsruhe Institute of Technology (KIT)},
    addressline={PO Box 3640},
    city={Karlsruhe},
    postcode={76021},
    country={Germany}
}

\affiliation[rug]{
    organization={Kapteyn Astronomical Institute, University of Groningen (RUG)},
    addressline={Landleven 12},
    city={Groningen},
    postcode={9747 AD},
    country={The Netherlands}
}

\begin{abstract}
    The radio detection technique of cosmic ray air showers has gained renewed interest in the last two decades. While the radio experiments are very cost-effective to deploy, the Monte-Carlo simulations required to analyse the data are computationally expensive. Here we present a proof of concept for a novel way to synthesise the radio emission from extensive air showers in simulations. It is a hybrid approach which uses a single microscopic Monte-Carlo simulation, called the origin shower, to generate the radio emission from a target shower with a different longitudinal evolution, primary particle type and energy. The method employs semi-analytical relations which only depend on the shower parameters to transform the radio signals in the simulated antennas. We apply this method to vertical air showers with energies ranging from $10^{17} \eV$ to $10^{19} \eV$ and compare the results with CoREAS simulations in two frequency bands, namely the broad [20, 500] \textMHz\ band and a more narrow one at [30, 80] \textMHz . We gauge the synthesis quality using the maximal amplitude and energy fluence contained in the signal. We observe that the quality depends primarily on the difference in \textXmax\ between the origin and target shower. After applying a linear bias correction, we find that for a shift in \textXmax\ of less than 150 \textgcm , template synthesis has a bias of less than 2\% and a scatter up to 6\%, both in amplitude, on the broad frequency range. On the restricted [30, 80] \textMHz\ range the bias is similar, but the spread on amplitude drops down to 3\%. These fluctuations are on the same level as the intrinsic scatter we observe in Monte-Carlo ensembles. We therefore surmise the observed scatter in amplitude to originate from intrinsic shower fluctuations we do not explicitly account for in template synthesis.
\end{abstract}

\end{frontmatter}

\section{Simulating radio emission from extensive showers}

Detecting cosmic rays using radio waves is a technique heavily researched in the 1960s and 1970s (see for example \cite{Jelley1965, Allan1971}). It was, however, abandoned in favour of other methods, which were able to extract the shower development more reliably. In the past two decades the field has experienced a revival thanks to the increased computational power and modern digital signal processing techniques \cite{Falcke2004, Huege2016}. Today, the radio technique has fully matured up to a point where it can be used alongside existing detection methods.

A unique advantage of the radio detection method is the low cost associated with the deployment of the experiments. This makes it easy to deploy the large-scale arrays required to detect the highest energy cosmic rays, the flux of which drops to fewer than one particle per $\text{km}^2$ per century \cite{Engel2011}. Instead of measuring the particles directly, we observe the cascade of secondaries they produce upon interacting in the atmosphere, in an event referred to as an extensive air shower (EAS). 

These EAS produce impulsive radio emission over a broad frequency range. It has been observed at frequencies ranging from $2 \MHz$ all the way up to $3 \GHz$ \cite{ALLAN1970, Smida2013}. Below 100 \textMHz\ the emission is coherent for antennas within a couple of 100m from the shower axis, meaning the power scales with the square of the number of particles present in the cascade and as such with the square of energy of the primary cosmic ray \cite{Huege2016}. On top of this, it has been shown that using radio detection techniques we are able to extract the atmospheric depth at which the EAS has the highest number of particles, denoted as the depth at shower maximum \textXmax\ \cite{LOPES2012, Buitink2014, Pont2021}.

Still, there are hints that there is much more to be learned from radio data \cite{Mitra2021, Corstanje2022}. Two interesting parameters are the width and asymmetry of the particle number longitudinal evolution, as they could help us distinguish between the predictions of different hadronic interaction models and even determine the proton fraction in the total primary flux \cite{Buitink2022}. 

Unfortunately, analyses are hindered by the requirement of performing a large number of simulations. Current state-of-the-art Monte-Carlo-based simulation codes calculating radio emission from air showers, such as CoREAS \cite{Heck1998, Huege2013} and ZHAireS \cite{Sciutto1999, Zhaires2012}, are computationally very expensive. The time required to run these microscopic models puts stringent limitations on the number of simulations that can be generated for each observation. Models based on macroscopic quantities exist, such as for example MGMR3D \cite{Scholten2008}. While these are much faster and are able to reconstruct basic shower parameters such as \textXmax\ \cite{Mitra2023}, it has still to be shown they can be used for general precision analyses.

In this work we develop the first step towards a novel simulation approach called \textit{template synthesis}. It is a hybrid model which characterises the radio signal in a single antenna using a parameterised function for the emission. The goal is then to find expressions describing the relation between the parameters of this function and the EAS properties. These semi-analytical expressions are extracted from a set of microscopic simulations, thus benefiting from their precision. If we have obtained these across different primary types, -energies and shower geometries, we will only ever need to run a single microscopic simulation for a certain geometry in order to synthesise the emission for showers with arbitrary longitudinal evolution profiles with the same geometry. 

Here we perform a simulation study using 600 vertical air showers simulated with CORSIKA together with its plugin CoREAS for the radio emission \cite{Heck1998, Huege2013}. From this simulation set we extract scaling relations that depend only on the \textXmax\ of the shower. Other dependencies are taken care of by rescaling the emission using the longitudinal profile. In this process we make use of air shower universality \cite{Lafebre2009, Lipari2009, Nerling2006, Gora2006, Giller2005}, which states that showers in the same stage of development have similar electron and positron distributions.

The rest of this work is organised as follows. First in Section \ref{sec:tools} we introduce the tools and conventions used in template synthesis. In Section \ref{sec:model} we present the template synthesis method. Next in Section \ref{sec:synthesis} we apply the method to our simulation set of vertical EAS and synthesise radio pulses filtered over two different frequency bands. The first one, 20 to 500 \textMHz, is the range we target for the application of our approach. Next to this, we investigate the 30 to 80 \textMHz\ band, which represents a typical range for current cosmic-ray radio experiments. We then present metrics to gauge the accuracy of template synthesis in Section \ref{sec:results}. In Section \ref{sec:discussion} we highlight some notable results and physical interpretations coming from the template synthesis approach. We conclude in Section \ref{sec:conclusion} with some final observations and a roadmap for the further development of the method.

\section{Aspects of radio emission from EAS} \label{sec:tools}

Throughout this work we will use the longitudinal profile $N(X)$ to refer to the number of electrons and positrons in the shower. In order to calculate the \textXmax\ of a particular air shower, we fit the longitudinal profile with the Gaisser-Hillas function as provided by CORSIKA \cite{GH77}. We do not consider other variations of the profile explicitly, such as for example the $L$ and $R$ parameters encoding the width and asymmetry of the profiles, respectively \cite{Andringa2011}. Still, for the synthesis procedure we do use the complete longitudinal profile, which encodes this information.

Hence, in template synthesis two showers with the same \textXmax\ but different $L$ or $R$ are only distinguished based on the absolute number of particles. However the change in slope of the longitudinal profile could also have an impact on the radio emission. Other fluctuations such as local over- and underdensities, and fluctuations in energy distributions are also not accounted for. All these aspects will introduce deviations, which in this work we will refer to as ``shower fluctuations''.

\subsection{Sliced simulations}

\begin{figure}
    \centering
    \includegraphics[width=0.8\linewidth]{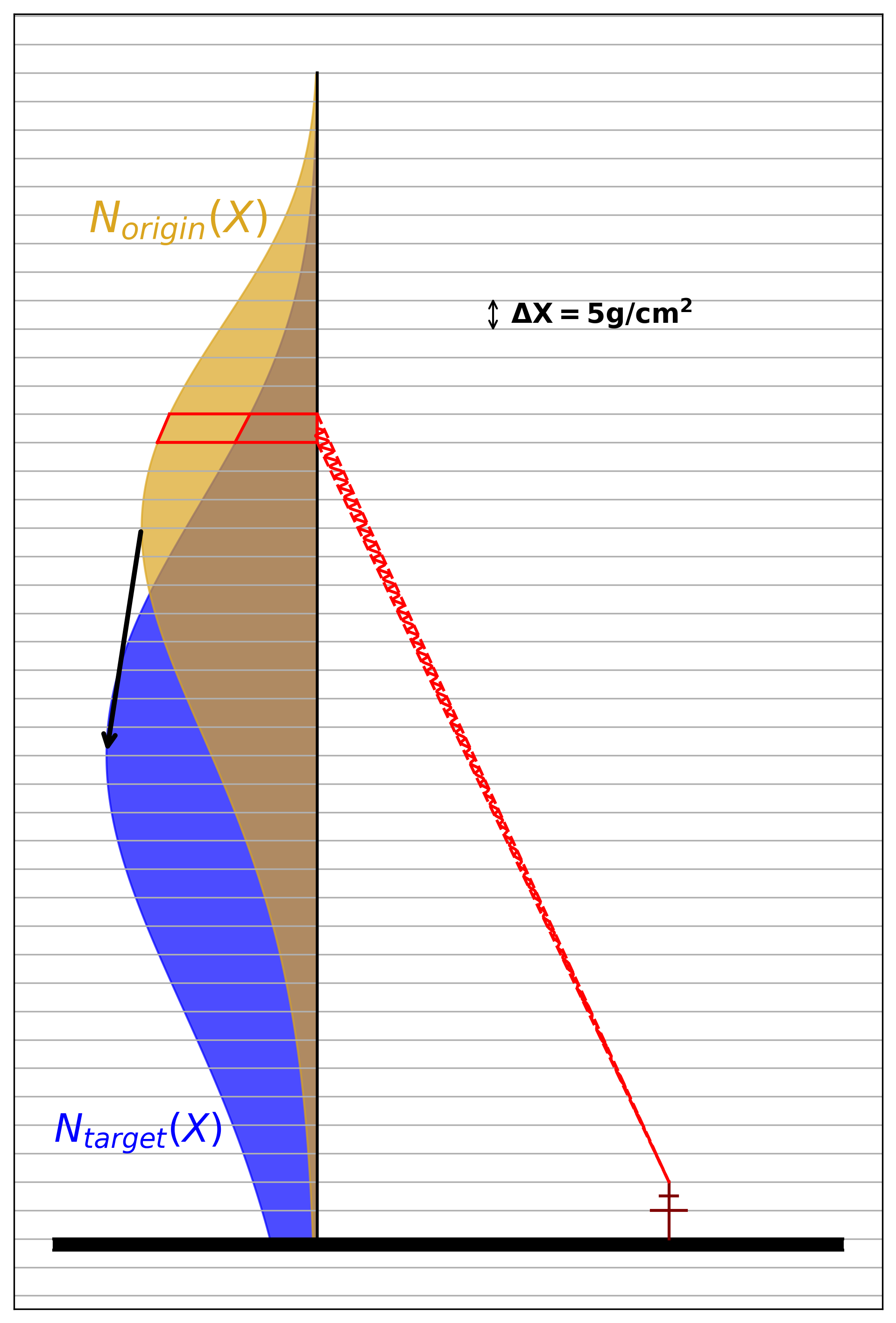}
    \caption{The atmospheric slicing performed for template synthesis. The vertical dimension represents the depth in the atmosphere, with the bold horizontal line being ground. In our simulations the antennas were simulated along the $\vec{v} \times \left( \vec{v} \times \vec{B} \right)$ axis. When scaling the emission from $N_{\text{origin}}$ to $N_{\text{target}}$, both the number of particles in the slice as well as the shower age are taken into account.}
    \label{fig:atm_slice}
\end{figure}

At the heart of the template synthesis technique, is the concept of ``sliced simulations''. These are simulated showers where we divided the atmosphere into layers of constant atmospheric depth\footnote{The technical implementation in CoREAS allows to have an arbitrary number of slices with arbitrary atmospheric depth, but we opt for a constant value.}, as shown in Figure \ref{fig:atm_slice}. This allows us to consider the radio emission coming from each slice $\vec{E}_{\text{slice}}$ separately, while still being able to reconstruct the complete signal $\vec{E} ( \vec{r}, t )$ in an antenna at position $\vec{r}$, by summing all the slices as

\begin{equation} \label{eq:summing}
    \vec{E} ( \vec{r}, t ) = \sum_{X} \vec{E}_{\text{slice}} ( X, \vec{r}, t ) \; .
\end{equation}
Here $X$ represent atmospheric depth in \textgcm. In CoREAS, the emission from a particle is counted towards a slice if the particle starts its track within that slice\footnote{Note that this differs from the particle counts in CORSIKA, which counts particles by placing virtual detector plans at specified atmospheric depths. As such a particle crossing multiple slices, would be counted in each of them. The radio emission however is only attributed to the slice it starts in.}.

Besides the emission in each slice, we also have the number of particles recorded in each of them. This information is crucial in order to synthesise across primary energies and particle types, as this number can vary dramatically between them. Thus from a sliced simulation we have both the radio emission $\vec{E}_{\slice}$ and the total number of electrons and positrons $N_{\slice}$ in each slice. 

\subsection{Emission mechanisms}

We know the radio emission from EAS is mostly generated by two independent mechanisms \cite{Huege2016}. The dominant one in air is due to the time-varying transverse currents arising from the deflection of the charged particles in the geomagnetic field in combination with their friction with atmospheric molecules, hence the name \textit{geomagnetic} emission. The other one is recognised as the Askaryan-effect, which occurs because of a time-varying excess of negative particles at the shower front. Therefore this is often referred to as \textit{charge-excess} emission. 

As the geomagnetic and charge-excess emission have a different linear polarisation, it is possible to decouple them \cite{Glaser2016}. With template synthesis we will synthesise both components separately. For this we will calculate the amplitude frequency spectrum $A_{\text{slice}} (f)$, 
\begin{align} \label{eq:dft}
    A_{\text{slice}}(f) = \biggl\lvert \sum_{n=0}^{N} s(n \cdot \Delta t) &\exp( -i 2 \pi \cdot n \cdot f \cdot \Delta t) \biggr\rvert \\
    \nonumber &f \in \left[ 0, \frac{1}{T}, \cdots, \frac{1}{2 \Delta t} \right] \;,
\end{align}
with $s(t)$ the time trace sampled at intervals $\Delta t$ and $T$ the duration of the time signal, for both components in every slice.

\subsection{Particle rescaling}

\begin{figure}
    \centering
    \includegraphics[width=\linewidth]{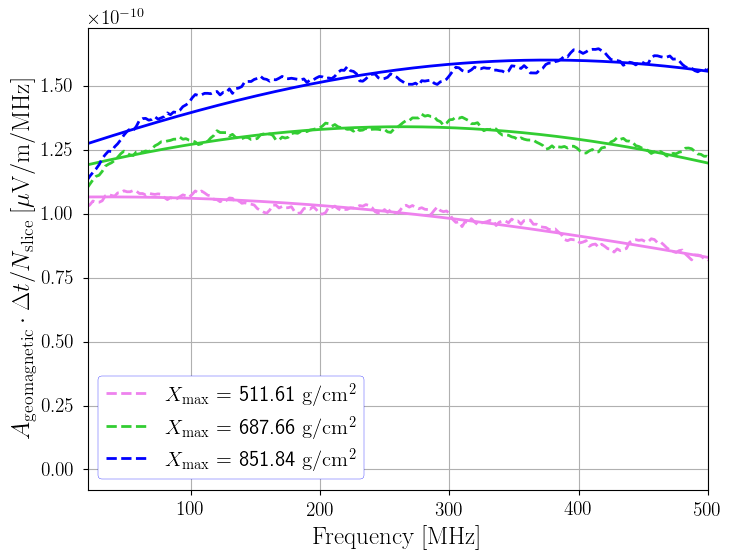}
    \caption{The amplitude frequency spectra for the geomagnetic component of three showers in the slice at 600 \textgcm\, in an antenna at 75m from the shower core. With the spectral rescaling approach, we fit a parabola in log-space to each of these, as shown by the solid line. The spectral parameters of this function will be used to capture how the spectra in a single slice depend on the \textXmax\ of the shower (a proxy for shower age in a particular slice). For this one slice, we can already see that the peak amplitude of the spectrum increases with \textXmax , i.e., with decreasing shower age.}
    \label{fig:fitted_spectra}
\end{figure}

The number of particles in a given slice determines to first order the amplitude of the radio emission from that slice. This number depends strongly on the primary energy of the cosmic ray. We could thus simply rescale the radiation coming from an input, origin shower in each slice with the particle number, yielding the synthesised electric field

\begin{equation} \label{eq:particle_rescaling}
    \vec{E}^{\text{synth}} (\vec{r}, t) = \sum_{X} \frac{N^{\target}(X)}{N^{\origin}(X)} \vec{E}^{\text{origin}}(\vec{r}, t, X) \; .
\end{equation}
However, as shown in \cite{Butler2019} this only works well for antennas far away from the shower axis. This approach does not generalise, because it turns out that the pulse shape of each slice depends on the stage of shower development (or shower age) in that slice. This can be demonstrated by comparing the emission from a fixed slice generated by three showers with different \textXmax\ (see Figure \ref{fig:fitted_spectra}). This is likely related to the energy distribution of the particles, which evolves to lower average energies as the shower ages. Hence the particle number is not sufficient information to know how the radio emission changes in each slice.

\subsection{Intrinsic variations in the radio emission} \label{subsec:variations}

A fundamental assumption we make in the template synthesis method is that two showers with an identical longitudinal profile produce the same radio emission. However, we find that the aforementioned shower fluctuations will, in general, introduce deviations from this assumption. In other words, when comparing the radio emission from two showers with the same longitudinal profile, we retrieve slight differences. To quantify how much this fluctuation is, we perform a simulation study to identify pairs of showers with almost identical profiles and compare the radiation in the antennas using the same metrics we will use later on to benchmark the template synthesis method.

The first metric takes the ratio of the maxima of the time traces $s_1(t)$ and $s_2(t)$,
\begin{equation} \label{eq:peak_ratio}
	S_{\text{peak}}(s_{\text{1}}, s_{\text{2}}) = \frac{\max(s_{\text{1}}(t))}{\max(s_{\text{2}}(t))} \; .
\end{equation}
Arguably a more important parameter in the context of modern reconstruction approaches, however, is the energy fluence contained in the pulses, defined as
\begin{equation}
	f(s) \propto \Delta t \sum_{n} s(n \cdot \Delta t)^2 \; ,
\end{equation}
with $n$ counting the samples in the trace and $\Delta t$ being the sample time. We then define a second metric which considers the ratio of the fluences,
\begin{equation} \label{eq:fluence_ratio}
	S_{\text{fluence}}(s_{\text{1}}, s_{\text{2}}) = \frac{f(s_{\text{1}})}{f(s_{\text{2}})} \; .
\end{equation}
Seeing fluence is an energy-like quantity, we need to half the uncertainties\footnote{We chose this approach instead of doing it the other way around, as the uncertainty on the amplitude is the more useful quantity when reproducing time traces.} coming from $S_{\text{fluence}}$ in order to compare them to the ones coming from $S_{\text{peak}}$. 

In order to find a set of pairs of similar air showers, we first performed 10,000 CONEX simulations. We then fitted each longitudinal profile with our own Gaisser-Hillas function, in the R-L formulation,
\begin{align}
    N(X) = \exp \left( \frac{\Xmax - X}{L \cdot R} \right)& \\ 
    \times & \left( 1 + \frac{R \cdot (X - \Xmax)}{L} \right)^{\frac{1}{R^2}} \; .\nonumber
\end{align}
For reference we calculate the 5\% and 95\% quantiles for the three parameters over the whole set. These are [552 , 777] \textgcm\ for \textXmax, [201 , 227] \textgcm\ for $L$ and lastly [0.16 , 0.36] for $R$. We then selected 60 pairs of showers which had an \textXmax\ and $L$ within 1 \textgcm\ and whose values for $R$ deviated no more than 0.01 . Then we re-simulated those 120 showers with CORSIKA together with the CoREAS plugin, for the same antenna set as described in Section \ref{sec:synthesis}.

In every antenna, we calculated the $S_{\text{peak}}$ and $S_{\text{fluence}}$ for each pair of similar showers. In order to better compare with the template synthesis benchmark later on, we calculate the quantities for the geomagnetic and charge-excess components separately. As the longitudinal profiles resulting from the CORSIKA simulation differ slightly from the ones which are simulated with CONEX, we normalise the traces with the total electromagnetic energy of the shower (calculated as the integral over the energy deposition by photons, electrons and positrons as reported by CORSIKA).

In Table \ref{tab:variations}, we report the mean and standard deviation of the 60 pairs in all antennas. Here we applied a bandpass filter from 20 \textMHz\ to 500 \textMHz\ to the traces. From this we gather that the shower fluctuations introduce on average 2-6\% fluctuations in the radio signal for this frequency band. The same analysis applied over the smaller [30, 80] \textMHz\ band yields fluctuations of the same order. Interestingly, the scatter is significantly larger for the charge excess component than the geomagnetic one. In the context of the template synthesis algorithm we present here, these values should be interpreted as a lower limit on the accuracy we can achieve. As we only use information about the particle number evolution, we cannot account for these other variations in the radio signal.

\begin{table*}
    \centering
    \begin{tabular}{|r|c|c|c|c|}
        \hline
        $r_{\text{ant}}$ & GEO $S_{\text{peak}}$ & GEO $S_{\text{fluence}}$ & CE $S_{\text{peak}}$  & CE $S_{\text{fluence}}$ \\
        \hline
        40m  & $1.000 \pm 0.020$ & $1.000 \pm 0.075$ & $1.002 \pm 0.048$ & $1.006 \pm 0.110$ \\
        75m  & $0.999 \pm 0.024$ & $0.998 \pm 0.083$ & $0.999 \pm 0.029$ & $1.001 \pm 0.088$ \\
        110m & $0.999 \pm 0.028$ & $0.999 \pm 0.089$ & $0.996 \pm 0.033$ & $0.994 \pm 0.093$ \\
        150m & $0.998 \pm 0.030$ & $0.999 \pm 0.093$ & $0.995 \pm 0.036$ & $0.992 \pm 0.098$ \\
        \hline
    \end{tabular}
    \caption{The mean and standard deviation of the two scoring metrics, as a function of the antenna distance from shower axis  $r_{\text{ant}}$, calculated over pairs of similar antennas (as defined in the main text). While calculating the metrics we filtered the traces in the [20, 500] \textMHz\ frequency band. This shows that shower fluctuations beyond the longitudinal profile introduce 2 to 6 percent fluctuations in the radio signal.}
    \label{tab:variations}
\end{table*}

\section{Template synthesis method} \label{sec:model}

The goal of template synthesis is to take the emission from a single microscopically simulated shower and use it to synthesise the emission from a shower with different parameters, such as primary particle, primary energy and longitudinal profile. In order to achieve this, we need scaling relations to inform us about the dependency of the emission on the shower parameters.

As we saw in the previous Section and Figure \ref{fig:fitted_spectra}, the frequency amplitude spectra in an atmospheric slice depend on both the primary energy and the shower age in that slice. Hence we consider these factors explicitly in our method in order to synthesise the radio emission.

Note that we only treat the amplitude spectra explicitly in template synthesis. For each slice we assume the phase spectrum does not depend on shower age or particle number.

\subsection{Parameterising amplitude spectra}

With template synthesis we synthesise the geomagnetic and charge-excess components separately. As we simulated our antennas on the $\vec{v} \times ( \vec{v} \times \vec{B} )$ axis, these can be retrieved by simply decomposing the trace into the $\vec{v} \times \vec{B}$ and $\vec{v} \times (\vec{v} \times \vec{B})$ polarisations, respectively.

We then parameterise both of them with a function $\tilde{A}_{\text{slice}} (f)$. For the geomagnetic component we use a parabola in log-space,

\begin{align} \label{eq:amplitude_geo}
    \nonumber \tilde{A}_{\text{geo}} (f,  X_{\slice}) & = \left( a_{\text{geo}} \cdot N_{\slice} \right)  \\
    \nonumber & \; \: \: \cdot \exp \left( b_{\text{geo}} \cdot (f - f_0) + c_{\text{geo}} \cdot (f - f_0)^2 \right) & \\
    & \quad \, + d \; ,
\end{align}
while for the charge-excess contribution we use
\begin{equation} \label{eq:amplitude_ce}
    \tilde{A}_{\text{ce}} (f,  X_{\slice}) = \left( a_{\text{ce}} \cdot N_{\slice} \right) \cdot \exp (b_{\text{ce}} \cdot (f - f_0)) + d \; ,
\end{equation}
i.e. we set $c = 0$, following \cite{Martinelli2022}. Here $f$ denotes frequency and $N_{\slice}$ the particle number in the slice. The explicit dependence on $X_{\slice}$ also indicates this is done for each slice separately.

Because the signal has no DC component, we introduce a shift in the frequency domain with $f_0$ in order to better constrain $a$. We call the $a$, $b$, $c$ parameters the \textit{spectral coefficients}. They depend on the slice and antenna position, as well as the shower properties such as \textXmax. Lastly $d$ represents a noise floor, either due to particle thinning in the Monte-Carlo simulations or to ``particle noise'', and is parametrised as

\begin{eqnarray}
    \nonumber
    \sqrt{d} = \text{max} 
    \left[ 
        10^{-9} \cdot \left( \frac{X_{\slice}}{400 \gcm} - 1.5 \right) 
    \right. \\
    \left. 
        \cdot \exp \left( 1 - \frac{r_{\ant}}{400 \m} \right), 0
    \right]
    \; ,
\end{eqnarray}
where $X_{\slice}$ is the atmospheric depth at the bottom of the slice and $r_{\ant}$ the distance from the antenna to the shower axis. This was tuned by eye, in order to match the incoherent noise floor across all simulated lateral distances.

\subsection{Extracting spectral functions}

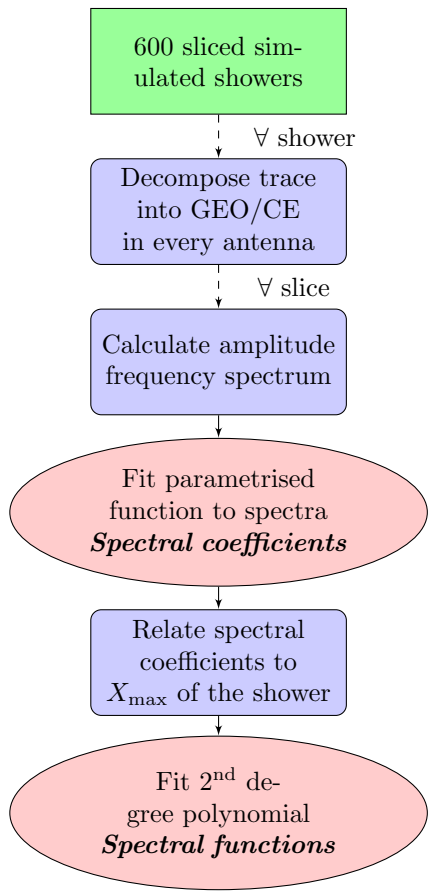
\begin{figure}
    \centering
    \begin{tikzpicture}[node distance = 2cm, auto]
        \node [start] (set) {600 sliced simulated showers};
        \node [below of=set, xshift=1.15cm, yshift=1cm] (text) {$\forall$ shower};
        
        \node [block, below of=set, node distance=2cm] (init) {Decompose trace into GEO/CE in every antenna};
        \node [below of=init, xshift=1cm, yshift=1cm] (text) {$\forall$ slice};
        
        \node [block, below of=init, node distance=2cm] (identify) 
            {Calculate amplitude frequency spectrum};
        \node [cloud, below of=identify, node distance=2cm] (spectra) 
            {Fit parametrised function to spectra \\ \textit{\textbf{Spectral coefficients}}};
        
        \node [block, below of=spectra] (stop) {Relate spectral coefficients to $X_{\text{max}}$ of the shower};
        \node [cloud, below of=stop] (fit) {Fit $2^{\text{nd}}$ degree polynomial \\ \textit{\textbf{Spectral functions}}};
        \path [line, dashed] (set) -- (init);
        \path [line, dashed] (init) -- (identify);
        \path [line] (identify) -- (spectra);
        \path [line] (spectra) -- (stop);
        \path [line] (stop) -- (fit);
    \end{tikzpicture}
    \caption{Schematic overview of how we extract the spectral functions from our simulation set. In every antenna, we decompose the emission from every slice in the geomagnetic (GEO) and charge-excess (CE) components. These are then fitted using Equations \eqref{eq:amplitude_geo} and \eqref{eq:amplitude_ce} respectively. Finally, we fit a parabola to each spectral parameter as a function of \textXmax. The result of this are the spectral functions.}
    \label{fig:flow_analysis}
\end{figure}

The shape of the amplitude spectrum $\tilde{A}_{\text{slice}} (f)$ will depend on \textXmax, as shown in Figure \ref{fig:fitted_spectra} and hence so will the spectral coefficients. We can fit this dependency of the coefficients in order to obtain the \textit{spectral functions}
\begin{align*}
    & a(r_{\ant}, X_{\slice}, \Xmax) \; , \\
    & b(r_{\ant}, X_{\slice}, \Xmax) \; , \\
    & c(r_{\ant}, X_{\slice}, \Xmax) \; ,
\end{align*}
in every slice. This procedure is applied to each antenna independently, indicated by the explicit dependency on the antenna distance. Therefore in the current version of template synthesis $r_{\ant}$, as well as $X_{\slice}$, can only take values on a fixed grid. We note here that while we write the functions as both a function of $X_{\slice}$ and \textXmax , a more accurate description should probably take some combination of the two that could serve as a proxy for shower age in the slice. We come back to this in Section \ref{sec:discussion}. 

We fit the spectral parameters as a function of \textXmax\ using a parabola. In order to better deal with the large scatter for some slices, especially the very early and late ones where there are only a few particles, we opt to first bin the data points by \textXmax . In each bin we calculate the mean value of the spectral parameter and the corresponding standard deviation. If a bin contains less than two data points, which can occur because not all showers might contain particles in that slice, we do not consider it for the parabolic fit. All the other bins are then fed to a least-squares fitting routine. We end up with a spectral function describing the spectral parameter value as function of the shower \textXmax\ in a given slice, for a particular antenna,
\begin{align*}
    & a(r_{\ant}, X_{\slice}, \Xmax) = p_0^a + p_1^a \cdot \Xmax + p_2^a \cdot \Xmax^2 \\
    & b(r_{\ant}, X_{\slice}, \Xmax) = p_0^b + p_1^b \cdot \Xmax + p_2^b \cdot \Xmax^2 \\
    & c(r_{\ant}, X_{\slice}, \Xmax) = p_0^c + p_1^c \cdot \Xmax + p_2^c \cdot \Xmax^2 \; .
\end{align*}

In Figure \ref{fig:flow_analysis} we present a schematic overview of how we extract the spectral functions. These need to be determined only once for the air shower geometry under consideration. Generalising them to arbitrary geometries will be the subject of future work.

\subsection{Construction of the template}

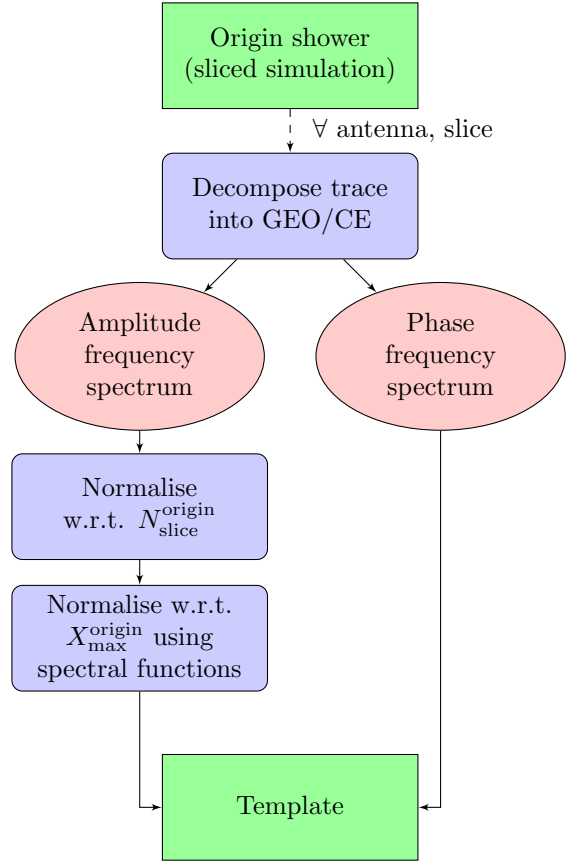
\begin{figure}
    \centering
    \begin{tikzpicture}[node distance = 2.5cm, auto]
        \node [start] (set) {Origin shower (sliced simulation)};
        \node [below of=set, xshift=1.5cm, yshift=1.5cm] (text) {$\forall$ antenna, slice};
        
        \node [block, below of=set, node distance=2cm] (init) {Decompose trace into GEO/CE};
        
        \node 
            [cloud, below of=init, xshift=-2cm, text width = 6em, node distance=2cm] (amp) 
            {Amplitude \\ frequency spectrum};
            
        \node 
            [cloud, below of=init, xshift=2cm, text width = 6em, node distance=2cm] (phase) 
            {Phase \\ frequency spectrum};
        
        \node [block, below of=amp, node distance=2cm] (norm) {Normalise w.r.t. $N_{\text{slice}}^{\text{origin}}$};

        \node [block, below of=norm, node distance=1.75cm] (norm2) {Normalise w.r.t. $X_{\text{max}}^{\text{origin}}$ using \\ spectral functions};
        
        \node [start, below of=set, node distance=10cm] (template) {Template};
        
        \path [line, dashed] (set) -- (init);
        \path [line,] (init) -- (amp);
        \path [line,] (init) -- (phase);
        \path [line,] (amp) -- (norm);
        \path [line,] (norm) -- (norm2);
    
        \draw [line,] (norm2) |- (template);
        \draw [line,] (phase) |- (template);
    \end{tikzpicture}
    \caption{To create a template, we start from a single microscopic sliced simulation, which we refer to as the origin shower. In each antenna, we again decompose the emission from each slice in the geomagnetic (GEO) and charge-excess (CE) components. From each component, we calculate the amplitude and phase frequency spectrum. The amplitude spectrum is then normalised with the number of particles in the slice, as well as the values from the amplitude function calculated using the spectral functions. The result from this is stored as the template amplitude frequency spectrum, together with the phase frequency spectrum. The latter is not treated explicitly in template synthesis.}
    \label{fig:flow_template}
\end{figure}

The final ingredient of template synthesis, is the \textit{template} itself. With this object and the spectral functions, we have all the necessary information to synthesise the emission from an air shower with arbitrary longitudinal profile.

In order to construct our template, we use a single microscopic simulation called the \textit{origin} shower. The origin is a microscopic simulation, sliced using the same procedure as the simulation set that was used to extract the spectral functions. From the origin shower we calculate the amplitude frequency spectrum $A_{\text{origin}}$ and phase frequency spectrum $\phi_{\text{origin}}$ in each antenna and every slice, as shown in Figure \ref{fig:flow_template}. Using the spectral functions with the \textXmax\ of the origin shower, we can normalise these to obtain the spectra of the template,

\begin{align*}
	A_{\template} (r_{\ant}, f, X_{\slice}) 
	&= A_{\text{origin}}(r_{\ant}, f, X_{\slice}) \\
    & \qquad \cdot 
	\left[ 
	    \tilde{A} (r_{\ant}, f,  X_{\slice} \Xmax^{\text{origin}})
	\right]^{-1} \\
	\phi_{\template} (r_{ant}, f,  X_{\slice})
	&= \phi_{\text{origin}} (r_{\ant}, f,  X_{\slice}) \; .
\end{align*}
In this equation the parameterised amplitude frequency spectrum $\tilde{A} (f)$ also depends on \textXmax\ through the spectral coefficients, which are calculated using the fitted spectral functions. As a result, in the following the frequency $f$ is restricted to the range used to fit the spectral functions.

The template thus contains a normalised version of the origin shower. It acquires its phase frequency spectrum as is, but the origin amplitude frequency spectrum is corrected for the longitudinal evolution (particle number in each slice and the \textXmax\ of the origin). As such, the template carries over the shower fluctuations from the origin shower.

\subsection{Synthesising the radio emission from an EAS}

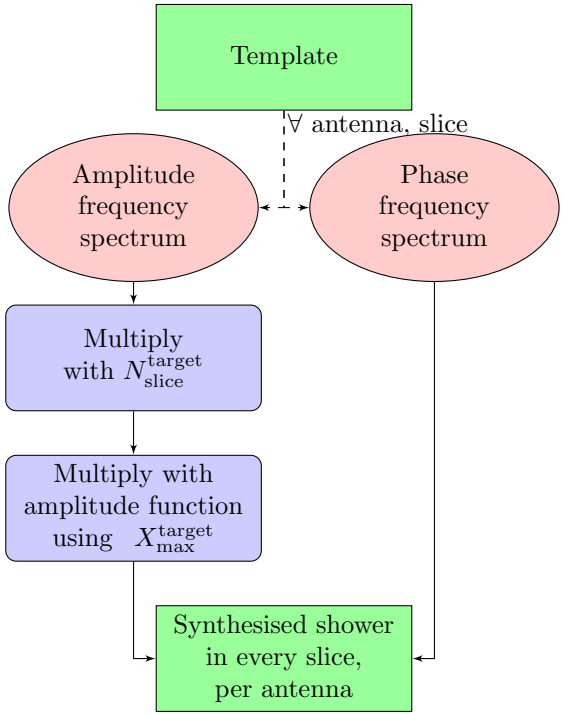
\begin{figure}
    \centering
    \begin{tikzpicture}[node distance = 2.5cm, auto]
        \node [start] (set) {Template};
        \node [below of=set, xshift=1.25cm, yshift=1.6cm] (text) {$\forall$ antenna, slice};
        
        \node 
            [cloud, below of=set, xshift=-2cm, text width = 6em, node distance=2cm] (amp) 
            {Amplitude \\ frequency spectrum};
            
        \node 
            [cloud, below of=set, xshift=2cm, text width = 6em, node distance=2cm] (phase) 
            {Phase \\ frequency spectrum};
        
        \node [block, below of=amp, node distance=2cm] (norm) {Multiply with $N_{\slice}^{\target}$};

        \node [block, below of=norm, node distance=2cm] (norm2) {Multiply with amplitude function using $\Xmax^{\target}$};
        
        \node [start, below of=set, node distance=8cm] (template) {Synthesised shower \\ in every slice, per antenna};
        
        \path [line, dashed] (set) |- (amp);
        \path [line, dashed] (set) |- (phase);
    
        \path [line,] (amp) -- (norm);
        \path [line,] (norm) -- (norm2);
    
        \draw [line,] (norm2) |- (template);
        \draw [line,] (phase) |- (template);
    \end{tikzpicture}
    \caption{To synthesise a shower with an arbitrary longitudinal profile, we start from the template. This object contains a (normalised) amplitude frequency spectrum and phase frequency spectrum, for every slice and in each antenna. The phase spectrum is again left untouched. To obtain the synthesised amplitude spectrum we multiply with number of particles in the slice, using the target profile, as well as the values of the amplitude functions using the target \textXmax . To obtain the complete synthesised signal in an antenna, we can sum up the contributions of all slices as per Equation \eqref{eq:summing}.}
    \label{fig:flow_synthesis}
\end{figure}

If we now take an arbitrary longitudinal profile, we can use the template together with the spectral functions to synthesise the amplitude spectrum over the selected frequency range in each slice. Together with the template's phase spectrum, we can retrieve the total radio signal in every antenna, as we show in Figure \ref{fig:flow_synthesis}. 

In each slice we can calculate the amplitude and phase spectra, given the particle number in that slice and the \textXmax\ of the target shower,
\begin{align*}
	A_{\target} (r_{\ant}, f, X_{\slice}) 
	&= A_{\template}(r_{\ant}, f,X_{\slice}) \\
    & \qquad \cdot \tilde{A} (r_{\ant}, f, X_{\slice}, \Xmax) \\
	\phi_{\target} (r_{\ant}, f, X_{\slice})
	&= \phi_{\template} (r_{\ant}, f, X_{\slice}) \; .
\end{align*}
These can then be recombined and transformed back to time-domain in order to retrieve the electric field traces in every slice. After summing all the slices per antenna as per Equation \eqref{eq:summing}, we obtain the complete synthesised signal in each antenna, bandpass filtered over the chosen frequency range.

\section{Spectral rescaling for the simulation set} \label{sec:synthesis}

Our simulation set is generated using CORSIKA v7.6400 with QGSJETII-04 and FLUKA as interaction models.  We apply thinning at a level of $10^{-7}$ with optimised weight limitation and put the low-energy cutoff for hadrons and muons at $0.3 \GeV$, and at $0.4 \MeV$ for electrons and photons. The magnetic field vector has a strength of 0.243 Gauss and an inclination of -35.7\textdegree . For the refractive index we used the Gladstone-Dale model with $n=1.000292$ at sea level.

In total our set contains 600 showers. One half of this set is initiated with a proton primary, while the other half has an iron nuclues as primary. For each primary, we simulate 100 showers with a primary energy of $10^{17} \eV$, another 100 showers at an energy of $10^{18} \eV$ and a last set of 100 showers with $10^{19} \eV$ as the primary energy. As mentioned before, all of them have vertical geometry (i.e. zenith angle $\theta = 0$) and we select the US standard atmosphere as parameterised by Keilhauer (atmospheric model 17 in the CORSIKA input file) \cite{Heck2023}.

The 4 simulated antennas are all taken to be at sea level on the $\vec{v} \times ( \vec{v} \times \vec{B})$ axis, at 40m, 75m, 110m and 150m from the shower axis. On this axis, the geomagnetic and charge-excess components have orthogonal polarisations, making it straightforward to decouple them. For the typical \textXmax\ in the considered energy range, the first antenna lies inside the Cherenkov ring, while the second one will be close to it. All other ones fall outside the ring.

Each simulation is sliced as described in Section \ref{sec:tools}. The slice thickness is taken to be 5 \textgcm\ , which is found to be a good middle ground between ensuring both sufficient population within each slice and a slice thickness which allows for realistic electromagnetic coherence \cite{Butler2020}. As such there are 207 slices in total, as the ground is at 1033.81 \textgcm\ for vertical showers. From now on, we will refer to slices by their upper limit in atmospheric depth. So the first slice, which runs from 0 \textgcm\ up to 5 \textgcm\ , will be described as the slice at 5 \textgcm\ .

\begin{figure*}
    \centering
    \includegraphics[width=\textwidth]{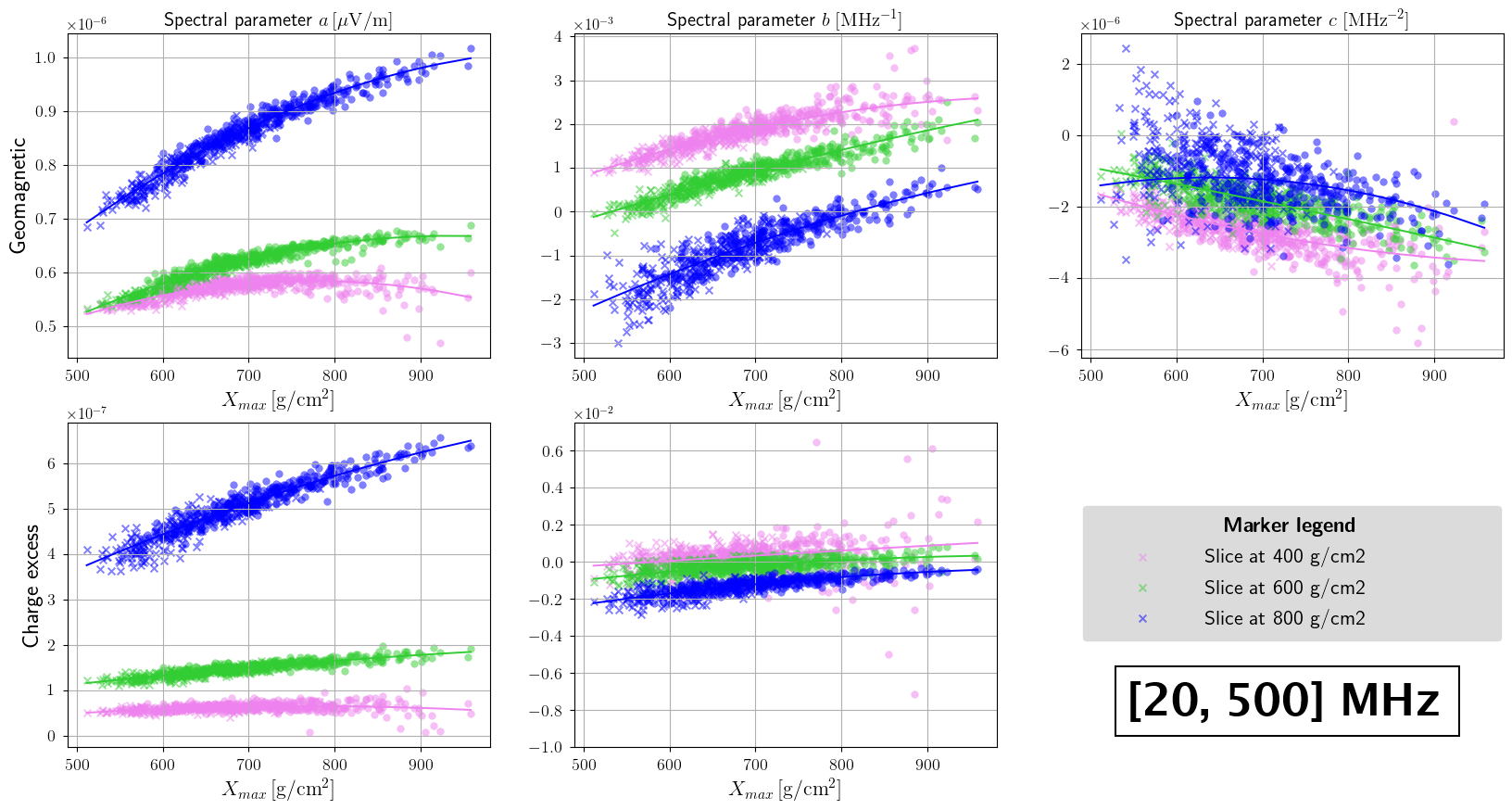}
    \caption{The values of the spectral parameters in three slices, indicated by the different colours, as a function of the \textXmax\ of the shower for the antenna at 75m from the shower axis. The fits were performed over the [20, 500] \textMHz\ frequency range. Showers with proton primaries are indicated with circles, while crosses indicate iron primaries. For each primary type the showers span primary energies from $10^{17} \eV$ to $10^{19} \eV$. Notice how all points line up on a single parabola, irrespective of primary particle type or energy. Overlayed in the same colour we also plotted the fit of this parabola.}
    \label{fig:spectral_fits_20_502} 
\end{figure*}

For each simulated shower, the above procedure generates 3 (vector components) times 4 (antennas) times 207 (slices) electric field traces, one for each slice in every antenna. We decompose every trace in its geomagnetic and charge-excess component, which in our setup amounts to taking the $y$ and $-x$ component (assuming the CORSIKA coordinate system, where $x$ points North and $y$ West) of the trace, respectively. Then we calculate the amplitude spectrum $A (f)$, with $f$ in \textMHz\, by applying Equation \eqref{eq:dft} to each component and fit the result using Equations \eqref{eq:amplitude_geo} and \eqref{eq:amplitude_ce}, respectively. If a particular slice contains no electrons or positrons, we exclude it from further analysis.

One important parameter in the fitting procedure is the frequency range over which to fit the spectrum. While a larger range would be more desirable as it would make the synthesis approach more widely applicable, it also results in less stable fits. This is because there are certain features over this broader range which our functional dependence cannot capture. In this work, we investigate two different bands of interest. The first one is [30, 80] \textMHz\ , used by multiple experiments such as LOFAR \citep{Schellart2013}, AERA \cite{Fuchs2012} and the upcoming Auger Radio Detector \cite{Castellina2019}. In addition, we investigate the broader frequency range of [20, 500] \textMHz\ which practically covers all the frequency bands used by current and most planned observatories. The figures presented in this work will use the latter frequency range, but where applicable performance metrics and results of the former band will be mentioned in the text.

In our analysis we set $f_0 = 50 \MHz$ for both frequency bands. We chose this as we can expect there to always be a reasonably strong signal around this frequency. At lower frequencies there is a sharp drop in amplitude, especially for antennas further away from the shower axis. At higher frequencies the signals starts to be dominated by numerical noise and particle thinning artefacts.  

In each antenna and slice, we investigate the relationship between the value of the spectral parameters and the \textXmax\ of the shower it was fitted to. As shown in Figure \ref{fig:spectral_fits_20_502}, for every slice the data points appear to follow a parabolic shape independent of the primary particle type or energy. The factor $N_{\slice}$ in Equations \eqref{eq:amplitude_geo} and \eqref{eq:amplitude_ce} is crucial to achieve this behaviour, as it removes the scaling of the amplitude with the primary energy. Furthermore, from air shower universality we know that proton and iron showers look very similar once their longitudinal profiles are rescaled to the same maximum value. Hence, once we take this out there should be no significant difference between the two, which is confirmed in Figure \ref{fig:spectral_fits_20_502} where the showers induced by proton and iron nuclei are marked with crosses and circles, respectively.

As described in Section \ref{sec:model}, we fit these relations with a parabola, obtaining the spectral functions. We note that these fits become somewhat arbitrary for slices with a low number of particles. This is clear in for example the top right panel of Figure \ref{fig:spectral_fits_20_502}, where we see that the $c$ parameter in slice 800 \textgcm\ exhibits a large scatter. Still, because of the low number of particles the contribution to the total radio signal is small. As such the impact of these lesser quality fits is minimal, as we can also see when quantifying the synthesis quality in Section \ref{sec:results}.

These spectral functions depend on the frequency range we consider. We compared the spectral functions for the two frequency bands we considered and noted similar values for the $a$ parameter. This is to be expected, as we fixed $f_0$ to the same value in both cases. The scatter on the $a$ parameter appears to be reduced for the smaller frequency band, but this comes at the cost of increased scatter on the other spectral parameter, $b$. This is probably due to certain features that  Equations \eqref{eq:amplitude_geo} and \eqref{eq:amplitude_ce} cannot describe properly. Although again we note that if we look at the synthesis quality in Section \ref{sec:results}, this does not seem to be an issue.

\section{Results and performance} \label{sec:results}

\begin{figure*}
    \centering
    \includegraphics[width=\textwidth]{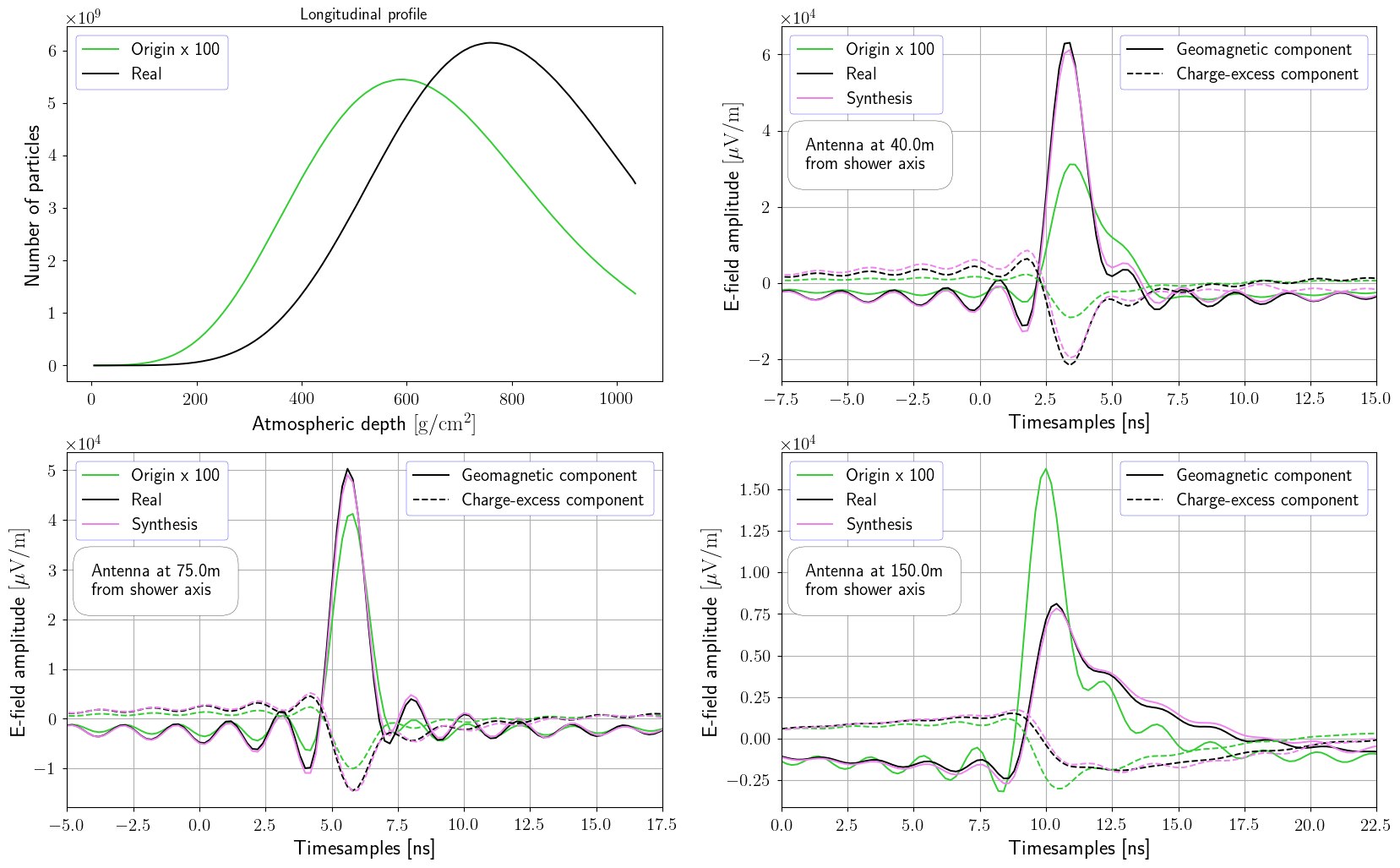}
    \caption{The top left panel shows the longitudinal profiles of the origin (green) and target (black) showers. The origin shower has an \textXmax\ of 589.87 \textgcm\ and was induced by a $10^{17} \eV$ iron primary, while the target shower was initiated by a $10^{19} \eV$ proton primary and has an \textXmax\ of 760.99 \textgcm.
    The other panels show the result of template synthesis using spectral rescaling in several antennas. The full and dashed lines are the geomagnetic and charge-excess components, respectively. In green we show the pulse from the origin shower and in black that of the target shower, as simulated with CoREAS. Overlayed in violet is the synthesised signal. 
    We multiplied the origin components by a factor of 100 for clarity, as to remove the first order rescaling with shower energy. All pulses are filtered over the [20, 500] \textMHz\ range.
    }
    \label{fig:synthesis_with_template}
\end{figure*}

In Figure \ref{fig:synthesis_with_template} we show the results of template synthesis for the [20, 500] \textMHz\ frequency range. In green we show the electric field traces of the origin shower, separated over the two components as indicated by the full and dashed lines. The primary of the origin shower was a proton with an energy of $10^{17} \eV$. As a target shower, we use the longitudinal profile of a CoREAS simulation from our simulation set. Hence we can extract its true radio signal from the simulation. The amplitudes of the origin shower are much smaller than those of the target shower, shown in black, because the target primary particle was a $10^{19} \eV$ proton. The method can easily handle this and produces pulses which are well-aligned with the CoREAS simulations.

\subsection{Performance metrics}

To get a better handle on the synthesis quality, we score the synthesis quality using the metrics defined in Section \ref{subsec:variations}. We compare the synthesised traces $s_{\text{synth}}(t)$ for a given target longitudinal profile with those coming from the underlying CoREAS simulation, $s_{\text{real}}(t)$. We gauge the performance of template synthesis over our simulation set by successively taking each of the 600 simulated showers as the origin and using it to synthesise the pulses for all other shower in the set. Due to the subsequent fitting and averaging of the data, the effects of a single shower are washed out enough that the method should not be biased towards it. Every time we score the synthesis quality using the two metrics defined above.

From this it becomes clear that one of the dominant factors in determining the synthesis quality is the difference between the \textXmax\ of the origin shower and that of the target shower. We observed this relation to not depend on values of the origin and target \textXmax , hence in the following we will consider the score to be solely a function of $\Delta X_{\text{max}} = \Xmax^{\text{target}} - \Xmax^{\text{origin}}$. 

In Figure \ref{fig:xmax_scoring} we plot the scores as a function of this variable. We clearly see that the further away template synthesis has to extrapolate the signals, the more inaccurate it becomes. We therefore opt to restrict the \textXmax\ range to which an origin shower can be synthesised. For the frequency range [20, 500] \textMHz\ we set the limit to 
\begin{align*}
    -200 \gcm < \Delta X_{\text{max}} < 150 \gcm \; ,
\end{align*}
while for the [30, 80] \textMHz\ this range can be extended to
\begin{align*}
    -250 \gcm < \Delta X_{\text{max}} < 150 \gcm \; .
\end{align*}

The reason for using asymmetric intervals, is that we observe the method performing worse when going to higher (i.e. positive) $\Delta X_{\text{max}}$. It is not immediately clear why the method should exhibit this asymmetry. Possible factors could be the change in atmospheric density, where the synthesis works better in one direction than the other. Another could be the change in distance from source to antenna, where near-field effects could cause complications. Lastly, during the synthesis process both the shape and the values of the amplitude frequency spectra need to be altered. When we go from a low to high \textXmax\ shower, we are transforming the emission coming from a slice in the tail of the origin longitudinal distribution to one before the target \textXmax . In this process, template synthesis needs to alter the spectral shape significantly as well as the particle number, which might blow up any inaccuracies in the synthesis.

When looking at the standard deviations (i.e. the dashed lines in Figure \ref{fig:xmax_scoring}), we see they are rather constant and in particular do not decrease significantly close to $\Delta X_{\text{max}} = 0$. This indicates we are dominated by the shower fluctuations not captured by our method, as described in Section \ref{subsec:variations}.

\subsection{Linear correction for $\Delta X_{max}$}

\begin{figure*}
    \centering
    \subfloat[Geomagnetic component with peak ratio]{
        \includegraphics[width=0.45\linewidth]{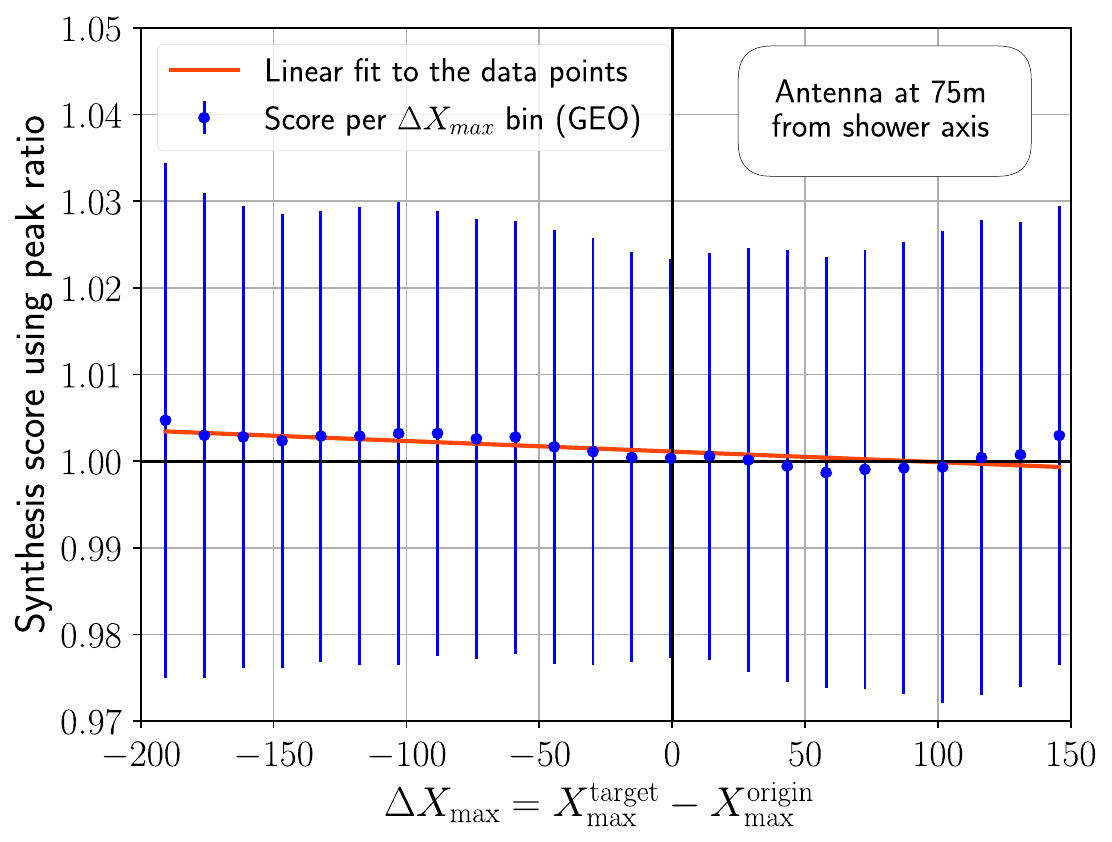}
    }
    \subfloat[Geomagnetic component with peak ratio, corrected using peak ratio linear fit]{
        \includegraphics[width=0.47\linewidth]{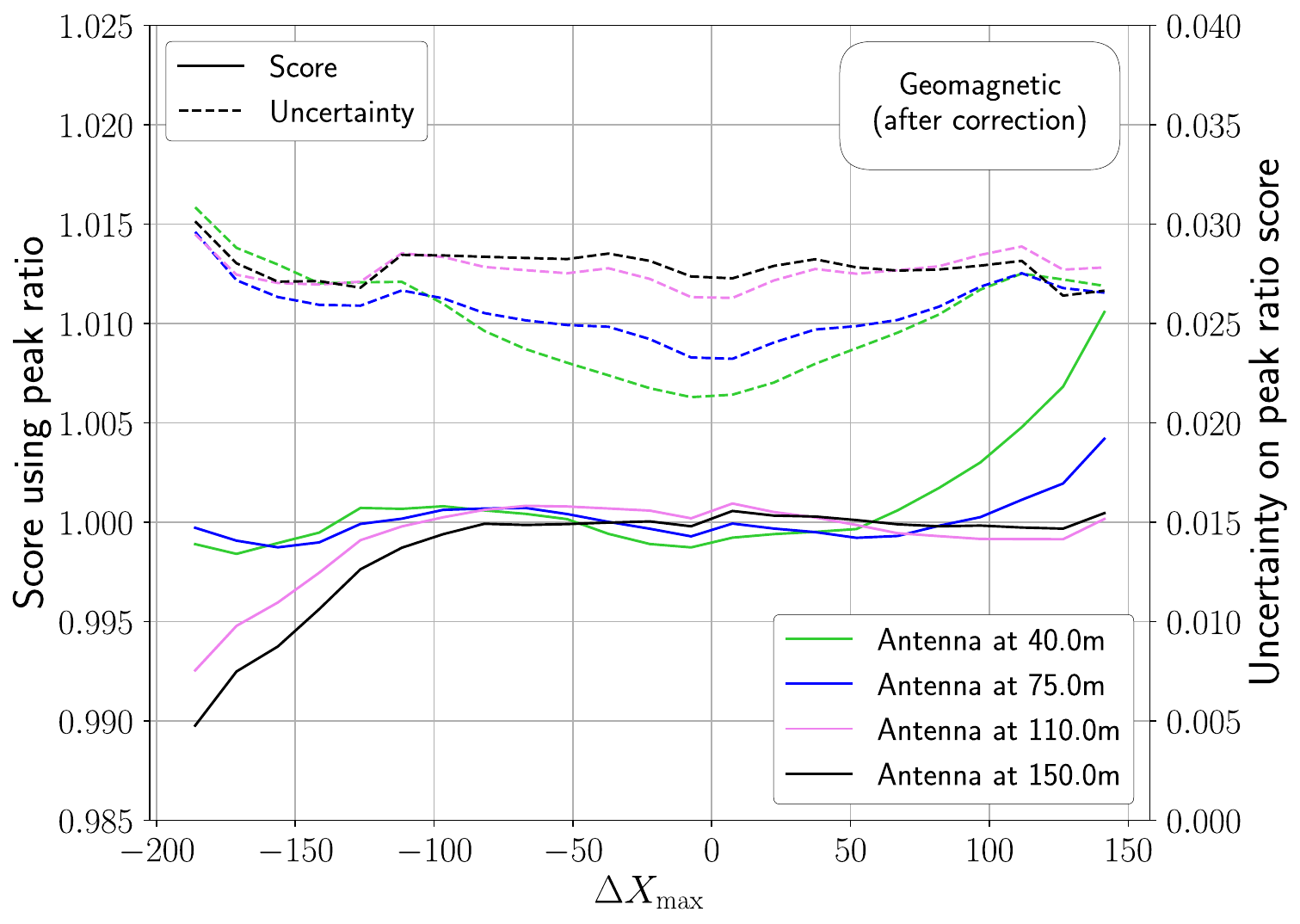}
    } \\
    \subfloat[Geomagnetic component with fluence ratio]{
        \includegraphics[width=0.45\linewidth]{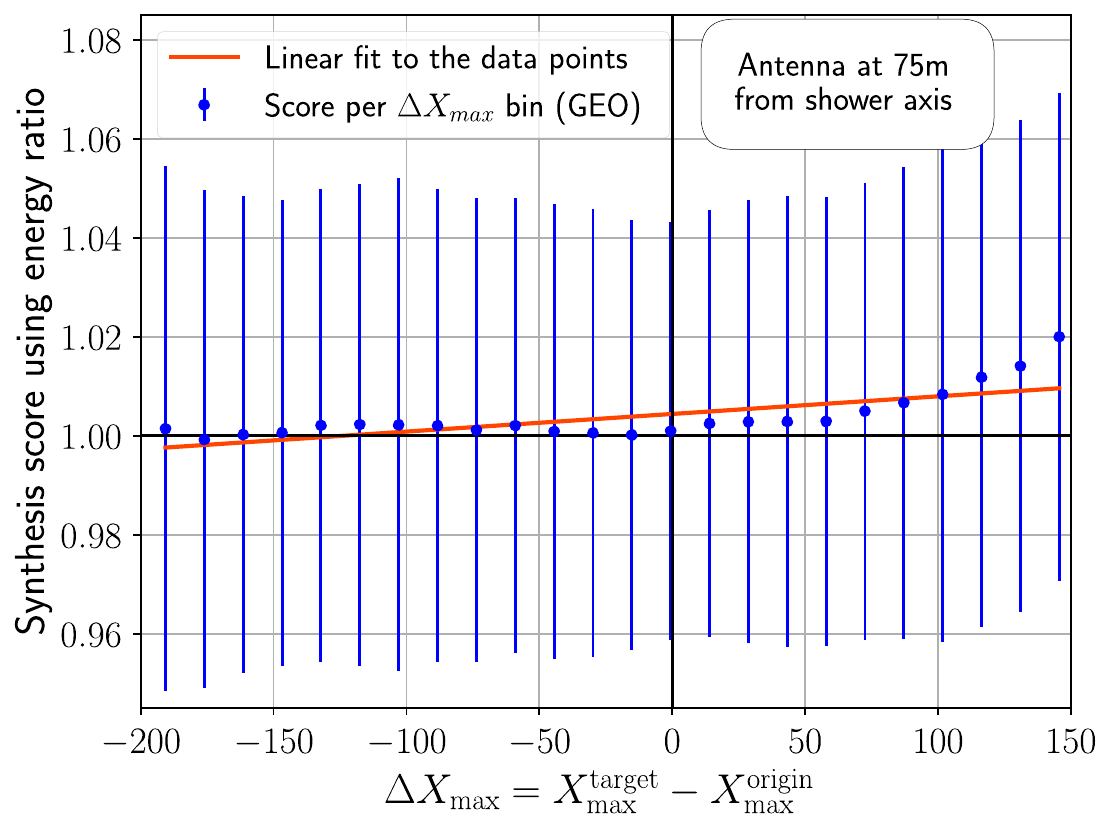}
    }
    \subfloat[Geomagnetic component with fluence ratio, corrected using fluence ratio linear fit]{
        \includegraphics[width=0.47\linewidth]{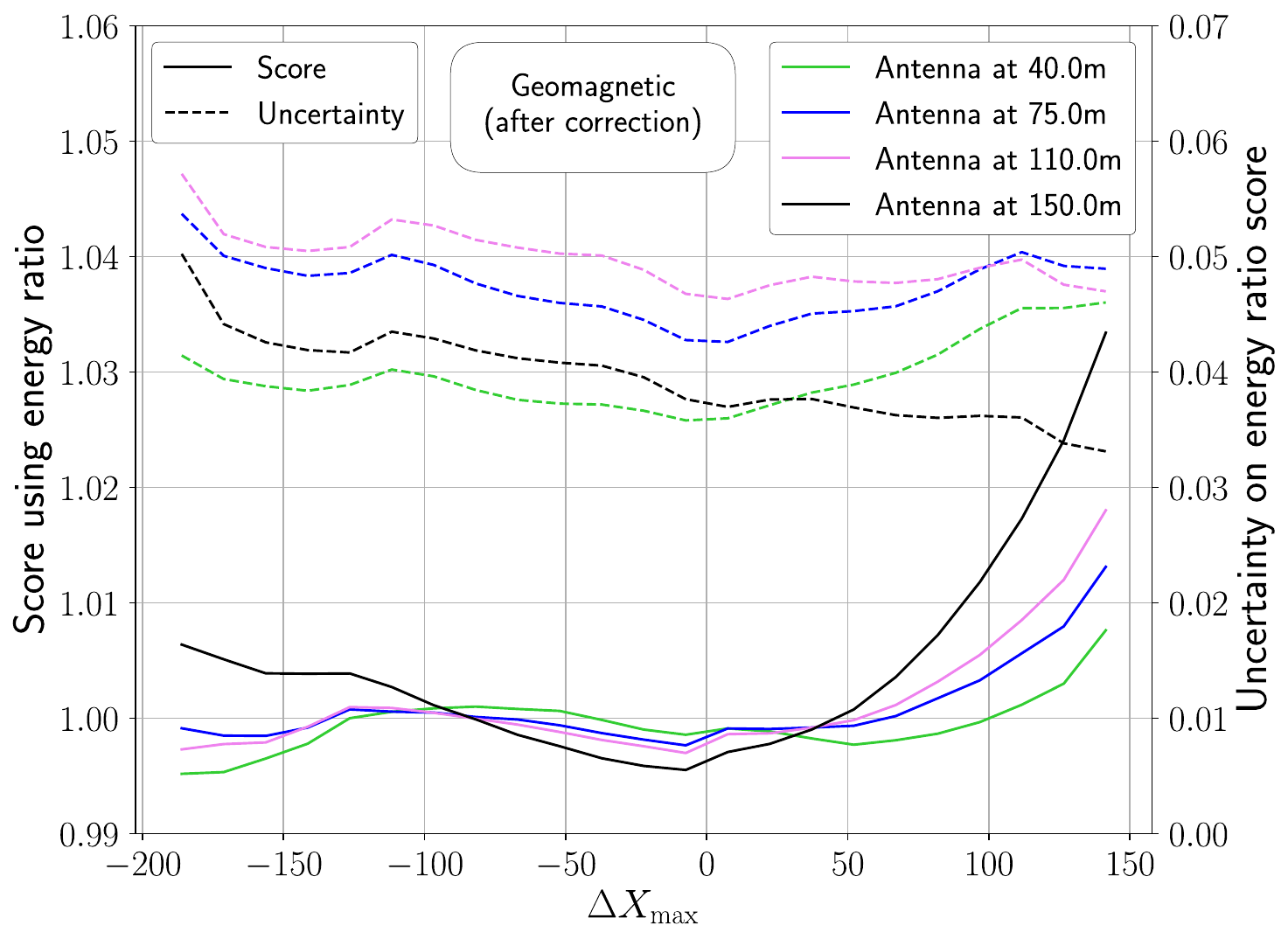}
    }
    \caption{The evolution of the scores for the geomagnetic component with $\Delta X_{\text{max}}$, in the [20, 500] \textMHz\ band. In panels (a) and (c) we show the mean and standard deviation of the scores for the antenna at 75m from the shower axis, binned in intervals of 15 \textgcm, for the peak ratio and fluence ratio scoring metrics, respectively. The red line represents a linear fit to the bias present in the data. We then remove this bias by correcting the synthesis result with this fit. The result is shown in panels (b) and (d), where we now show the mean (full lines) and standard deviation (dashed lines) for all antennas.}
    \label{fig:xmax_scoring}
\end{figure*}

Taking a closer look at how the scores evolve with $\Delta X_{\text{max}}$ in Figure \ref{fig:xmax_scoring} (a) and (c), we notice the method has a $\Delta X_{\text{max}}$-dependent bias which to first order can be described by a linear function. This bias gets more pronounced as we move away from the Cherenkov cone (not shown here). We defer a more thorough analysis of this effect to later work, when we will include other geometries than purely vertical air showers as well.

Here we opt to improve template synthesis for the vertical case, by fitting a straight line to each score binned in $\Delta X_{\text{max}}$ for every antenna and component separately. This gives an estimate of the bias the method has according to the different metrics, which we can factor out during the synthesis process by multiplying the synthesised electric field amplitudes with the inverse of the bias. The biggest correction that gets applied this way is 3.5\% following the peak ratio scores and 4.6\% according to the fluence ratio metric.

After correction, the method retains a slight bias up to a maximum of 1.7\% in amplitude at the highest $\Delta X_{\text{max}}$ for the geomagnetic component, as shown in Figure \ref{fig:xmax_scoring} (b) and (d). Still, this bias is smaller than the overall spread, which is around 3\% according to $S_{\text{peak}}$ and around 5\% following $S_{\text{fluence}}$ for the [30, 80] \textMHz\ band. Again, seeing that $S_{\text{fluence}}$ is an energy-like quantity this translates into a 2.5\% spread on the amplitudes. These are on the same level as the intrinsic variations we found in Section \ref{subsec:variations}, so we suspect the method is limited by not using information beyond the longitudinal profile in this frequency band.

For the [20, 500] \textMHz\ band we find an accuracy of 6\% and 10\% for $S_{\text{peak}}$ and $S_{\text{fluence}}$ respectively, yielding an average 5-6\% spread on the amplitude. We note that for this band the scores are much better for the geomagnetic component than for the charge-excess contribution. For the latter, we observe for several showers mismatches between the real and synthesised signals far away from the main pulse. This has a strong impact on the fluence ratio score, notably for large positive $\Delta X_{\text{max}}$, which is reflected by the worse accuracy. Whether this has a substantial impact in practical applications remains to be seen however, and will be investigated in later work.

For both frequency bands we can explain the observed scatter by the shower fluctuations we do not account for in our method, as they are not represented in the longitudinal profile. These can be interpreted as natural variations that occur when generating a shower with a desired \textXmax . Hence one would observe the same scatter when generating an ensemble of showers with all similar \textXmax\ using a Monte-Carlo approach. Therefore we do expect the accuracy of template synthesis to be good enough to be used in analyses.

\section{Discussion} \label{sec:discussion}

While the previous sections highlight the procedure with which we apply the template synthesis method, here we wish to delve deeper in some notable aspects we encountered.

First of all we want to come back to Figure \ref{fig:fitted_spectra} and the realisation that the pulse shape in each slice depends on the \textXmax\ of the shower. Additionally, within a slice the spectra from different showers arrange themselves in a clear hierarchy according to the \textXmax\ value. We surmise this can be understood from the perspective of shower age. The \textXmax\ in a given slice, or more accurately the \textit{difference} between the atmospheric depth of the slice and \textXmax\ of the shower, is a proxy for the age of the shower in that slice. The age of a shower represents the state of its development in the atmosphere, and as such the state of the currents responsible for the radio emission. Several properties of the air shower have been shown to correlate strongly with shower age, such as energy, angular and lateral distributions of electrons and positrons \cite{Lafebre2009, Lipari2009}.

For this reason, we presume we can simplify the problem further. In the current version of template synthesis we consider the spectral functions to be unique for every slice, allowing for maximal complexity in the method. Given that shower age is the determining factor for pulse shape, it should be possible to reformulate the spectral functions as
\begin{align*}
    & a(r_{\ant}, X_{\slice} - \Xmax) \; , \\
    & b(r_{\ant}, X_{\slice} - \Xmax) \; , \\
    & c(r_{\ant}, X_{\slice} - \Xmax) \; .
\end{align*}
Then, the spectral functions are no longer specific to a slice, which reduces the number of parameters to be saved significantly. Furthermore, this formulations lends itself much better to generalising towards and across other, non-vertical geometries. 

This also highlights the fact that we do not rely on the shower profiles to behave like Gaisser-Hillas distributions. As we input the number of particles in every slice, it should in principle be possible to synthesise the emission from unusual shower profiles, such as the so-called ``double-dump'' showers. The only ambiguity in this case would be the choice of what \textXmax\ describes the shower age in the slices. Following the current procedure, the fit would probably select an \textXmax\ somewhere between the two peaks, which should describe some kind of average shower age. It will be interesting to compare this approach to superposing the radio emission from two showers which do have a Gaisser-Hillas-like longitudinal profile, and hence more better defined \textXmax , and whose sum gives the double-bump profile.

In order to generalise the method to arbitrary geometries, the most obvious approach would be to explore the dependency of the spectral functions on the air shower geometry. Given the success for the vertical case, the method should be applicable to other fixed geometries without much modification. Indeed, we have recently shown this works for air showers with a 45\textdegree\ zenith angle \cite{Desmet2023}. The next step thus seems to be to interpolate the spectral functions between the different geometries. An interesting question that arises here is how to relate the slices for different geometries, i.e. should one compare slices with the same physical distance or grammage to the ground.

Another avenue that could be explored would be to look at different parametrisations of the amplitude frequency spectrum. Also including other shower fluctuations explicitly, would add more dimensions in which template synthesis can work and thus probably improve the quality as well. It is even possible this will be required in order to obtain a more general formulation of the spectral functions, as we discussed above.

\begin{figure}
    \centering
    \includegraphics[width=\linewidth]{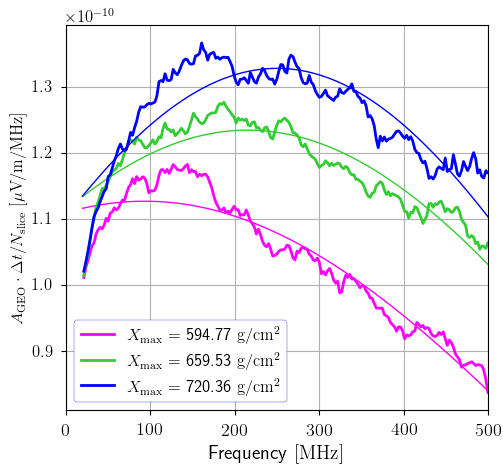}
    \caption{The amplitude frequency spectra from the geomagnetic component of the radio emission from 3 different showers, in the slice at 600 \textgcm . Even though our parametrised function (overlayed line of the same colour as the spectra) captures the general trend, some modulation on top of it is apparent. For example, at lower frequencies the parametrisation undershoots the geomagnetic spectra, while around 250 \textMHz\ the spectra start overshooting slightly. By rescaling the emission in every slice, we can retain these trends.}
    \label{fig:doublebump}
\end{figure}

In order to arrive at a better parametrisation, we will first need a better understanding of the coherency conditions of the radio emission. While on the Cherenkov cone we observe a strong signal across the whole frequency band, away from it the amplitude spectra can exhibit interesting patterns. Recently it has become clear these are not solely due to numerical noise in the Monte Carlo simulations, as similar effects have been observed in results from macroscopic codes. While our paramatrisation does capture the general trend for all amplitude spectra, for some combinations of slice and antenna we observe a modulation on top of it, as shown in Figure \ref{fig:doublebump}. This also explains why we do not use the parametrised amplitude frequency spectrum for the target shower. By rescaling the emission from an origin shower, we can preserve these more intricate features.

We are also aware of other approaches to reduce computational cost for simulations such as MGMR3D \cite{Scholten2008} and Radio-Morphing \cite{Zilles2020, Chiche2021}. The latter approach was developed specifically for very inclined air showers, for which the source of the radio emission is much further away than for vertical showers, simplifying the problem considerably. We expect template synthesis to converge towards Radio-Morphing as the zenith angle increases, in the sense that we will obtain similar scaling relations in that regime. We plan to test this convergence as well as compare template synthesis to the macroscopic approach of MGMR3D once template synthesis is capable of handling all geometries.

Lastly, we want to address the fact that template synthesis is only capable of synthesising signals at the antenna positions for which the spectral functions were extracted. To overcome this, we envision combining template synthesis with a pulse interpolation scheme, in order to provide an integrated solution to synthesise the radio emission from an EAS at any chosen antenna location. Several interpolation approaches have been explored. In \cite{Alvarez2014} the authors exploited the polarisation properties of the two dominant emission mechanisms. More recently a method based on Fourier interpolation has been implemented \cite{Corstanje2023}.

\section{Conclusion and outlook} \label{sec:conclusion}

We presented an initial proof of concept for a novel approach to synthesise the radio emission from EAS, called template synthesis. The method considers the radio emission in every antenna, coming from slices of constant atmospheric depth separately. It uses semi-analytical relations, which are extracted from a set of microscopic CoREAS simulations, to rescale the amplitude frequency spectra from each slice. At the end of the process the contributions from all slices are summed up to recover the complete signal. Other modifications, such as minor bias corrections, are also applied during the final step.

In this work we applied template synthesis to a set of 600 vertical air showers, simulated using CORSIKA with the CoREAS plugin, as a proof-of-principle. 
From this simulation set we extracted the semi-analytical relations by parameterising the amplitude frequency spectrum in each antenna and in every slice. The parameters of this function exhibit a quadratic dependence on the \textXmax\ of the shower. By fitting this dependence, we obtained the spectral functions needed to account for changes in the radio-emission pulse shape associated with the ageing of air showers and with which we could synthesise the radio emission from an air shower with an arbitrary longitudinal profile.

In order to quantify the performance of the method, we compared the results of template synthesis to the pulses simulated by CoREAS. We used every shower from our simulation set as the origin shower to synthesise the emission from all other showers in the set and scored the result using two different metrics. 
When doing so, we observed a bias which was linear with the shift in \textXmax\ from origin to target shower. As we only looked at one geometry here, we opted to not look deeper into this effect. Rather we fitted this bias with a linear function as function of $\Delta X_{\text{max}}$ and corrected the final result of template synthesis for it.

Throughout this paper, we considered two frequency bands. The first, [30, 80] \textMHz , is a band used by several of the leading cosmic ray radio experiments. The second one is [20, 500] \textMHz\ and covers all frequencies over which the radio pulse is expected to be coherent. Our results show that the smaller frequency band yields a better synthesis quality, with the average deviation in amplitude around 3\% down from 6\%. This is probably due to functional form with which we parametrise the amplitude frequency spectra, as it starts to break down over larger frequency intervals. Investigation of other functional forms is something we defer to later work, once we have applied template synthesis to other air shower geometries.

We showed that after bias correction, the method works well for vertical air showers, if the change in \textXmax\ between origin and target shower is not too large. For the frequency band of 30 to 80 \textMHz , the synthesis has an accuracy better than 3\% in amplitude if we restrict $-250 \gcm < \Delta X_{\text{max}} < 150 \gcm$.  For the broader frequency range, template synthesis achieves an accuracy better than 6\% in amplitude over the interval $-200 \gcm < \Delta X_{\text{max}} < 150 \gcm$. The poorer accuracy seems to be the result of deviations from the real trace away from the main pulse. The impact of this in practical applications will be the subject of further studies. 

These fluctuations are on the same level as the intrinsic scatter we observed in Monte-Carlo ensembles with the same longitudinal profile. These fluctuations will introduce scatter in the synthesis quality plots, as they differ in other aspects than \textXmax . Apparently, they originate from variations in other shower characteristics such as the energy distribution of the particles or the lateral density profile. When the template synthesis method is used to generate an ensemble of showers based on a single input shower, these natural variations are not included. Therefore, an uncertainty of a few percent needs to be assigned to reproduce the correct spread. Nonetheless, this implies we captured the variations we did account for well and that the method itself is working as intended for the vertical geometry we considered. New insights into the relation between shower age and the radio emission from the EAS, as well as the coherency conditions at the antennas, could be readily implemented in the method and reduce the scatter even further. Still, template synthesis does not seem to introduce an appreciable uncertainty on top of the natural scatter resulting from shower fluctuations. Hence it should not be the bottleneck when used for reconstruction analyses.

Our next goal is to extend template synthesis to cover all geometries. Given the success for fixed geometries we are left with the task to understand how the spectral functions scale with geometry, which will allow us to interpolate them. If we have the spectral functions for a given geometry, we would only need one microscopic origin shower to act as the origin, to synthesise the emission for any other shower with the same geometry.

\bibliographystyle{elsarticle-num} 
\bibliography{sources.bib}

\end{document}